\documentclass[preprint,aps,superscriptaddress]{revtex4-2}
\usepackage{amsmath}
\usepackage{amsfonts}
\usepackage{hyperref}
\usepackage{graphicx}
\usepackage{float}
\usepackage{dcolumn}
\usepackage{bm}
\usepackage{tabularx}

\usepackage{color,xcolor}

\def\be{\begin{equation}}
\def\ee{\end{equation}\noindent}
\def\bear{\begin{eqnarray}}
\def\ear{\end{eqnarray}\noindent}
\def\bec{\blue\begin{equation}}
\def\eec{\end{equation}\black\noindent}
\def\bearc{\blue\begin{eqnarray}}
\def\earc{\end{eqnarray}\black\noindent}
\def\benn{\begin{enumerate}}
\def\enn{\end{enumerate}}

\def\bk{\bf k}
\def\non{\nonumber\\}
\def\half{\frac{1}{2}}
\def\2F1{\phantom{}_2F_1}

\def\e{\,{\rm e}}

\def\bk{{\bf k}}

\def\b0{{\bf 0}}

\begin{document}

\title{Generalized Gelfand-Dikii equation and solitonic electric fields for fermionic Schwinger pair production}

\author{Naser Ahmadiniaz}
\email{n.ahmadiniaz@hzdr.de}
\affiliation{Helmholtz-Zentrum Dresden-Rossendorf, Bautzner Landstra\ss e 400, 01328 Dresden, Germany}
\author{Alexander M. Fedotov}
\email{al.m.fedotov@gmail.com}
\affiliation{Centro Internacional de Ciencias A.C., Campus UNAM-UAEM, 62100 Cuernavaca, Morelos, Mexico}
\author{Evgeny G. Gelfer}\email{egelfer@gmail.com}
\affiliation{ELI Beamlines facility, The Extreme Light Infrastructure ERIC, Dolni Brezany 252 41, Czech Republic, }
\author{Sang Pyo Kim}
\email{sangkim@kunsan.ac.kr}
\affiliation{Department of Physics, Kunsan National University, Kunsan 54150, Korea}
\affiliation{Center for Relativistic Laser Science, Institute for Basic Science, Gwangju 61005, Korea}
\author{Christian Schubert}
\email{christianschubert137@gmail.com}
\affiliation{Centro Internacional de Ciencias A.C., Campus UNAM-UAEM, 62100 Cuernavaca, Morelos, Mexico}

\begin{abstract}
In previous work on Schwinger pair creation in purely time-dependent fields, 
it was shown how to construct ``solitonic'' electric fields that do not create scalar pairs with an arbitrary fixed momentum.
We show that this construction can be adapted to the fermionic case
in two inequivalent ways, both closely related to supersymmetric quantum mechanics for reflectionless potentials, and
both leading to the vanishing of the density of created pairs at certain values of the
P\"oschl-Teller like index $p$ of the associated Schr\"odinger equation. For one of them, we are able to demonstrate that the pair non-creation
can be interpreted as a quantum interference effect using the phase-integral formalism.
Asymptotically for large $p$, here scalar particles are not created for integer $p$ and fermions are not created for half integer $p$.
Thus for any given momentum
we can construct electric fields that create scalar particles but not spinor particles, and vice versa.
In the scalar QED case, the solitonic fields had originally been found using the Gelfand-Dikii equation, which is related to the resolvent of
the mode equation, and through it to the (generalized) KdV equation \cite{geldik}. This motivates us to develop for the spinor QED case, too,
an evolution equation that can be considered as a fermionic generalization of the Gelfand-Dikii equation. 
\end{abstract}


\maketitle

\section{Introduction}\label{intro}

The creation of pairs from vacuum by an electromagnetic or gravitational background field is a nonperturbative quantum effect that probes the quantum structure of the vacuum. Schwinger pair production by an electromagnetic field \cite{Heisenberg:1935qt,Schwinger:1951nm} and Hawking radiation by a black hole \cite{Hawking:1974sw} are two of the most prominent consequences of quantum field theory.
The parametric interaction of a particle with the background field transforms the in-vacuum into the out-vacuum that
in general consists of multiparticle states of the in-vacuum, when a proper definition of the in- and out-vacua
can be made in two asymptotic regions (for a comprehensive review of the in-out formalism, see \cite{DeWitt:1975ys,DeWitt:book03}, and Schwinger pair production in QED \cite{Kim:2008yt}).
Experimentally,
the recent development of intense lasers based on chirped pulse amplification (CPA) has led to an intensity of $ 1.1 \times 10^{23}~\mathrm{W/cm^2}$ \cite{Yoon2021Realization},
the highest that has been reached for optical lasers so far,
and may open a window for Schwinger pair production in electromagnetic fields in the future \cite{DiPiazza:2011tq,fikks}.
However, there are very few electric field profiles for which the Klein-Gordon or Dirac equation are solvable in terms of known special functions,
so that it is convenient to have master equations, suitable for numerical evaluation, for the mean number of produced pairs due to generic electric fields
(see \cite{healgi2} and refs. therein).


An important limiting case is the one of an electric field that is purely time-dependent but spatially homogeneous.
This is because then for a given particle momentum the pair creation problem can be formulated in terms of an effective quantum mechanical evolution equation.
For the scalar QED case, a suitable master formula (“quantum Vlasov equation”) for the density of created pairs
has been known and used for decades \cite{kescm1,kescm2,klmoei,healgi,schton,sbrpst,bmprssv,ahrsv,dumlu,fedotov_prd2011}
\begin{eqnarray}
\dot{\cal N}_{\bf k} (t)
&=& \frac{\dot\omega_{\bf k} (t)}{2\omega_{\bf k}(t)} \int_{-\infty}^{t} dt' \, \frac{\dot\omega_{\bf k} (t')}{\omega_{\bf k}(t')}
(1+2 {\cal N}_{\bf k} (t') )  \cos \Bigl\lbrack 2\int_{t'}^{t} dt'' \omega_{\bf k} (t'') \Bigr\rbrack \, ,
\label{inout}
\end{eqnarray}
where $\omega_{\bf k}^2 (t)$ is the energy squared,
\bear
\omega_{\bf k}^2 (t) = \Pi_{\bf k}^2 + \mu_{\bf k}^2, \quad \Pi_{\bf k} = k_{\parallel} - qA_{\parallel}, \quad \mu_{\bf k}^2 = m^2 + {\bf k}_{\perp}^2\, .
\label{defomega}
\ear
Here and in the following we use the temporal gauge $A_0 = 0, \dot {\bf A}(t) = - {\bf E}(t)$.
The field points into a fixed direction, and the subscripts $\parallel, \perp$ refer to the components along the field respectively parallel and perpendicular to it.
${\cal N}_{\bf k}(t)$ is usually taken to be zero initially, i.e. at $t=-\infty$, and
for $t\to\infty$ turns into the density of created particles with fixed momentum $\bf k$.

At intermediate times, ${\cal N}_{\bf k}(t)$ is inherently ambiguous on account of its dependence on the choice of an instantaneous adiabatic basis \cite{dabdun2014,dabdun2016}. Its physical meaning is presently still under discussion \cite{fedotov_prd2011,dabdun2014,dabdun2016,gamamascal,gamamaspin,ilderton,gamamavlasov}.

The derivation of \eqref{inout} relies, in quantum field theory terms, on the in-out formalism in the Furry picture: the in-vacuum and the out-vacuum, and their Fock bases of annihilation and creation operators.
An equivalent but somewhat simpler evolution equation can be obtained using the in-in formalism
\cite{Kim:2011jw}.
In this case ${\cal N}_{\bf k}(t)$ at large times in general does not directly correspond to the density of created pairs, rather some asymptotic averaging is to be performed to obtain that quantity
\cite{Huet:2014mta}.

Coming to the physically more interesting case of fermion QED, in this context of pair creation by a time-dependent field there have been a number of studies exploring the 
dependence of the number of created pairs on spin \cite{mirpop-72,huta-19,alnasri-23,strxue-15,woba-15,kohlfuerst, cakiki}.
An evolution equation analogous to \eqref{inout} for fermion QED has been developed in \cite{schton,sbrpst} and was shown to be equivalent to other approaches in \cite{fedotov_prd2011,dumlu}.
It can be written as
\begin{eqnarray}
\dot{\cal N}_{\bf k} (t)
&=& \frac{\mu_\bk q \dot A_{\parallel}(t)}{2 \omega_{\bf k}^2 (t)}
 \int_{-\infty}^{t} dt' \,
 \frac{\mu_\bk q \dot A_{\parallel}(t')}{\omega_{\bf k}^2 (t')}
\bigl(1-2 {\cal N}_{\bf k} (t') \bigr)  \cos \Bigl\lbrack 2\int_{t'}^{t} dt'' \omega_{\bf k} (t'') \Bigr\rbrack
\, .
\label{inoutspin}
\end{eqnarray}
The evolution equations \eqref{inout}, \eqref{inoutspin} have turned out to be very suitable for brute-force numerical
evaluation, as well as for other approximation schemes \cite{healgi,sbrpst,bmprssv,ahrsv,dumlu,fedotov_prd2011,kaos}.
However, their structure as integro-differential equations makes them rather intractable for attempts at closed-form analytic evaluation. For analytic purposes, one would like to take advantage of the fact that the Hamiltonian of a charged scalar field in a time-dependent electric field is equivalent to an infinite system of modes that are just time-dependent harmonic oscillators.
This suggests the application of Lewis-Riesenfeld theory \cite{lewrie}, and indeed
this theory can be used to determine the exact quantum states
of the oscillators in terms of quantum invariants, as shown explicitly in \cite{Kim:2011jw}.

Nevertheless,
experience has shown that it is often preferable to switch from the mode equation to some other equivalent equation. In particular, it was shown in
\cite{dabdun2014,dabdun2016} that the density of created pairs can be expressed in terms of the solutions to the Gelfand-Dikii equation, a third order linear equation \cite{geldik} which easily incorporates the in-out formalism as well as the in-in formalism by properly choosing the boundary data in either the far past or future. One interesting aspect of the formulation in terms of the Gelfand-Dikii equation is that this equation relates to the resolvent of the mode equation, and through it to the (generalized) KdV equation \cite{geldik}.
In \cite{Kim:2011jw,Huet:2014mta,Kim:2021jxw} this connection was used to construct, starting from the well-known solitonic solutions of the KdV equation, an infinite family of electric fields that are tuned to {\sl not} pair create at a given (fixed but arbitrary) reference momentum $\tilde k$. Their vector potentials are given by
\begin{eqnarray}
qA_{\parallel(p)}(t) \equiv  \tilde k_{\parallel} - \sqrt{\tilde k_{\parallel}^2
+ \frac{p(p+1)\tilde\omega_0^2}{\cosh^2(\tilde\omega_0 t)}} \, ,
\label{Anscal}
\end{eqnarray}
so that the corresponding squared energies at $\bf k = \tilde {\bf k} $ are of P\"oschl-Teller type,
\bear
\omega^2_{(p){\bf k}}(t) = \omega_0^2 \Bigl(1+ \frac{p(p+1)}{\cosh^2 (\omega_0 t)}\Bigr) \, .
\label{omegapt}
\ear
Here we have introduced
\bear
\omega_0 \equiv \sqrt{{\bf k}^2+m^2}\,,
\ear
and similarly $\tilde \omega_0$. Using the well-known connection between the density of created pairs and
the reflection probability of the associated over-the-barrier scattering problem (see \cite{dumdun-interference} and refs. therein)
one concludes that pair creation is absent for $\bf k = \tilde {\bf k}$ whenever the
reflection coefficient is zero, which is the case when the P\"oschl-Teller index $p$ takes integer values
$p=1,2,\ldots$ (see, e.g., \cite{flugge}, problem 39). Thus, although the total density of created pairs does
not vanish (there are good reasons to believe that for a purely time-dependent electric field
this can never happen \cite{63}) there appears a tuneable zero in the spectrum of created pairs.
This might be seen as a first step towards the construction of what has been called ``designer fields'' by
F. Hebenstreit \cite{kmvha,hebenstreit}.
See \cite{kolosh} for a similar application of reflectionless potentials and the underlying supersymmetry to produce surprising null effects of external fields
in atomic physics. In QFT, such ``pair non-creation'' has already been studied in various contexts (see \cite{Vachaspati:2022ayz} and refs. therein).

The asymptotic vanishing of the density of created pairs for these solitonic examples also shows in the most drastic possible way that
a non-zero density of created pairs at finite times should not yet be interpreted as indicative of the presence of particles.

Let us emphasize that this vanishing is exact, which distinguishes this class of examples from applications of coherent quantum interference
to Schwinger pair creation such as \cite{dumdun-interference,akkdun} where superpositions of alternating sign pulses are used to create interference patterns in momentum space,
but still relying on a small ${\cal N}_k$ approximation.

The present paper is devoted to the generalization of some of the results of \cite{Kim:2011jw,Huet:2014mta,Kim:2021jxw} to the spinor QED
case. In particular, we derive a generalization of the Gelfand-Dikii equation to the
spin-half case, and construct spinor QED analogues of the ``solitonic'' electric fields
\eqref{Anscal} in two different ways.

The organization of the paper is as follows. In Section \ref{mode_equation} we summarize various equivalent known ways of
encoding the time evolution of a scalar field interacting with a purely time-dependent electric field.
In Section \ref{sec:Scalar pair creation} we show for the scalar case  how to calculate the density of created pairs from a solution of the Gelfand-Dikii equation,
and we generalize that equation to the fermionic QED case in Section
\ref{sec:Generalization of the Gelfand-Dikii equation to the spinor QED case}.
In Section \ref{sec:Solitonic fields in scalar QED} we summarize previous work on the
solitonic backgrounds for the scalar case, explaining, in particular, how the pair non-creation comes about for integer
values of the P\"oschl-Teller index $p$. Here the exact solution of the mode equation is known and we compare it with numerical study of the Vlasov equation in (\ref{inout}), achieving a good numerical control both for pair creating and noncreating values of the P\"oschl-Teller index $p$. We then study the same backgrounds for the fermionic case in Section \ref{sec:Solitonic fields in spinor QED: the original solitons}, presenting numerical
evidence that pair non-creation persists for certain values of that index that now depend on the chosen reference momentum.
In Section \ref{sec:Solitonic fields in spinor QED: the alternative solitons} we introduce a new, alternative set of solitonic backgrounds, derive the corresponding criterion for non-creation
and verify it both analytically and numerically for a number of examples. Here the corresponding Schr\"odinger equation can be solved
exactly, and in Section \ref{sec:Solitonic pair non-creation as a Stokes phenomenon} we moreover show how to derive the pair non-creation as a quantum interference phenomenon.
In Section \ref{sec:Conclusions} we summarize our finding and discuss possible extensions.
In the appendix we derive the solutions of the scalar and spinor mode equations for the
Sauter field, of the scalar mode equation for the original solitons, and of the spinor mode equation for the alternative solitons.

\section{The mode equation and its many disguises}
\label{mode_equation}

The Hamiltonian of a charged field with mass $m$ and charge $q$ in a purely
time-dependent electric field can be decomposed into an
infinite number of harmonic oscillators with time-dependent frequencies
(see, e.g., \cite{ps}),
\begin{eqnarray}
H(t) = \sum_{\bf k} \pi_{\bf k}^* \pi_{\bf k} + \omega_{\bf k}^2 (t) \phi_{\bf k}^* \phi_{\bf k} \, , \label{ham}
\end{eqnarray}
where $\omega_{\bf k}^2 (t)$ has been defined in \eqref{defomega}.
We will generally assume that the electric field is localized in time, $E(\infty) = E(-\infty)=0$
with a finite integral $\int_{-\infty}^{\infty} dt E(t) = A(-\infty)-A(\infty)$.
It follows that we can define the limits
\bear
\omega_i \equiv \lim_{t\to - \infty} \omega_{\bf k}(t)\, , \quad
\omega_f\equiv \lim_{t\to  \infty} \omega_{\bf k}(t) \, .
\label{defomegapm}
\ear
The simplicity of the purely time-dependent electric field case is due to the fact that the
individual modes $\phi_{\bf k}$ obey a time-dependent oscillator equation (“mode equation”)
\begin{eqnarray}
\ddot{\phi}_{\bf k} (t) + \omega^2_{\bf k} (t) \phi_{\bf k} (t) = 0 \, .
\label{meqscal}
\end{eqnarray}
Canonical quantization of the scalar field together with the commutation condition for the scalar field creation and annihilation operators lead to the Wronskian constraint
%
\begin{eqnarray}
{\rm Wr} [\phi_{\bf k}, \phi^*_{\bf k}] \equiv   \phi_{\bf k}\dot\phi^*_{\bf k} - \dot\phi_{\bf k}\phi^*_{\bf k}=  i \, .
\label{wronskian}
\end{eqnarray}
From the mode equation \eqref{meqscal} together with the constraint equation \eqref{wronskian}
one can, by taking derivatives, derive the following evolution equation for $\vert \phi_{\bf k}\vert \equiv \rho_{\bf k}$,
\bear
\ddot \rho_{\bf k}(t) + \omega_{\bf k}^2(t) \rho_{\bf k} (t) = \frac{1}{4\rho_{\bf k}^3} \, ,
\label{ermileq}
\ear
which is called Ermakov-Milne equation \cite{ermakov,milne}.
It can be considered as equivalent to the mode equation since the information on the phase of $\phi_{\bf k}$ can be recovered by unitarity (see, e.g., \cite{dabdun2016}).
Neither is information lost working with either the real or the imaginary part of $\phi_\bk$ alone \cite{gamamascal,gamamaspin}.

Yet another evolution equation can be constructed passing from $\phi_{\bf k}$ to
$G_{\bf k} \equiv \vert \phi_{\bf k}\vert^2$.
This equation is a linear third-order differential equation,
\bear
\dddot{G}_{\bf k}  + 4 \omega^2_{\bf k}  \dot{G}_{\bf k}  + 4 \dot\omega_{\bf k} \omega_{\bf k}  G_{\bf k} = 0 \, .
\label{geldikeq}
\ear
It was derived in \cite{Kim:2011jw} in terms of the variable
\bear
F_{\bf k} = \frac{1}{\omega_i} \biggl(\frac{1}{2\omega_i} - G_{\bf k} \biggr) \, ,
\ear
which obscured the fact that \eqref{geldikeq} is actually the Gelfand-Dikii equation \cite{geldik}.
Let us collect here also a few useful facts about this equation. It can alternatively be written
as a non-linear second-order equation,
\bear
2G_{\bf k} \ddot G_{\bf k} - \dot G_{\bf k}^2 + 4\omega_{\bf k}^2 G_{\bf k}^2 = 1 \, ,
\label{GDs2}
\ear
(the first integral of \eqref{geldikeq}),
which leads to the following relation for any two independent solutions $G_{\bf k}^{(1)}$
and $G_{\bf k}^{(2)}$,
\bear
\frac{d}{dt} {\rm Wr}[\dot G_{\bf k}^{(1)}(t),\dot G_{\bf k}^{(2)}(t)] = 4 \dot\omega_{\bf k}(t)\omega_{\bf k} (t)
\,{\rm Wr}[G_{\bf k}^{(1)}(t),G_{\bf k}^{(2)}(t)] \, .
\label{wron}
\ear
Finally, defining 
\bear
\chi_{\bf k} \equiv \frac{1+i \dot{G}_{{\bf k}}}{2G_{\bf k}}\,,
\label{defchi}
\ear
Equation (\ref{GDs2}) can be transformed into a Ricatti equation:
\bear
i\dot\chi_{\bf k}  + \chi_{\bf k}^2 = \omega_{\bf k}^2 \, .
\label{chieq}
\ear
In fact, $\chi_{\bf k}=i\dot{\phi}_{\bf k}/\phi_{\bf k}$ recovers (\ref{meqscal}) and gives $\phi_{\bf k}=\sqrt{G_{\bf k}} \e^{-i\int\frac{dt}{2G_{\bf k}}}$ which satisfies the Wronskian constraint (\ref{wronskian}).
For a detailed discussion of the relation between these various evolution equations see \cite{dabdun2014}. 


\section{Scalar pair creation}
\label{sec:Scalar pair creation}

It is easily shown (see, e.g., \cite{fedotov_prd2011,Huet:2014mta}) that from a solution $\phi_\bk$ of the classical mode equation \eqref{meqscal} one can get a solution
\bear
{\cal N}_\bk(t) := \frac{|\dot \phi_\bk (t)|^2 + \omega_\bk^2(t) |\phi_\bk(t) |^2}{2\omega_\bk(t)} - \frac{1}{2}
\label{relNphi}
\ear
of the Vlasov equation (\ref{inout}).
The mean number (\ref{relNphi}) counts the energy of each mode measured by the instantaneous energy
$\omega_\bk(t)$ beyond the vacuum energy $\omega_\bk(t)/2$.
By differentiation of the mode equation \eqref{meqscal}, one can express $|\dot \phi_\bk (t)|^2$ in terms of
$G_\bk\equiv |\phi_\bk|^2$:
\bear
|\dot \phi_\bk|^2 = \omega_\bk^2 G_\bk + \frac{1}{2} \ddot G_\bk \, .
\label{elphidot}
\ear
This makes it also possible to express ${\cal N}_\bk(t) $ in terms of $G_\bk$,
\bear
{\cal N}_\bk(t) = \frac{\ddot G_\bk}{4\omega_\bk(t)} +\omega_\bk(t) G_\bk - \frac{1}{2} 
\label{NtoG}
\ear
Usually one assumes that initially there are no particles present, ${\rm lim}_{t\to -\infty}{\cal N}_\bk(t)=0$.
The relevant solution of the mode equation will then obey (up to an irrelevant phase factor)

\bear
\phi_\bk(t)\, \stackrel{t\to -\infty}{\longrightarrow}\, \frac{1}{\sqrt{2\omega_i}} \,\e^{-i\omega_i t} \, .
\label{condin}
\ear
For $G_\bk$ this implies that
\bear
{\rm lim}_{t\to -\infty} G_\bk(t) = \frac{1}{2\omega_i} \, .
\label{Ginitial}
\ear
This initial condition fixes the constant on the rhs of (\ref{GDs2}) since $\dot{G}_{\bf k}(-\infty)=\ddot{G}_{\bf k}(-\infty)=0$, and $\mathcal{N}_{\bf k}(-\infty)=0$.

\section{Generalization of the Gelfand-Dikii equation to the spinor QED case}
\label{sec:Generalization of the Gelfand-Dikii equation to the spinor QED case}

Proceeding to the case of spin $\frac{1}{2}$ particles and introducing the spinor field in the form $\Psi(t)=\psi^{(\pm)}_{\bf k} u_\pm$,
where  $\gamma^0\gamma^3 u_\pm=\pm u_\pm$, from the squared version of the Dirac equation we obtain the analogue of the mode equation (\ref{meqscal}) for the functions
$\psi^{(\pm)}_{\bf k}(t)$ \cite{basefr,schton,fedotov_prd2011}
\begin{eqnarray}
\ddot{\psi}_{\bf k}^{(\pm)} (t) + \bigl[\omega^2_{\bf k} (t) \mp iq \dot A_{\parallel}(t)\bigr]  \psi_{\bf k}^{(\pm)} (t) = 0
\, .
\label{meqspin}
\end{eqnarray}
The superscript $(\pm )$ refers to the projection of the spin on a fixed axis, which in the following will
be chosen as the field direction.
In  \cite{fedotov_prd2011} it was shown that the initial condition \eqref{condin} has to be
replaced by (in our present conventions)
\bear
\psi_{\bf k}^{(\pm)}(t)
\stackrel{t\to -\infty}{\longrightarrow}\, \frac{\sqrt{\omega_i\pm \Pi_{\bf k}(-\infty)}}{\sqrt{2\omega_i}} \,\e^{-i\omega_i t}
=
 \frac{\sqrt{\omega_i\pm \bigl(k_{\parallel}-qA_{\parallel}(-\infty)\bigr)}}{\sqrt{2\omega_i}} \,\e^{-i\omega_i t}
 \, ,
\label{condinferm}
\ear
that the constraint equation \eqref{wronskian} has to be replaced with
\bear
\omega_\bk^2 \vert \psi_{\bf k}^{(\pm)}\vert^2 + \vert\dot\psi_{\bf k}^{(\pm)}\vert^2 \mp i\Pi_{\bf k}
(\dot\psi_{\bf k}^{(\pm)} \psi_{\bf k}^{(\pm)\ast} - \psi_{\bf k}^{(\pm)}\dot\psi_{\bf k}^{(\pm)\ast}) = \mu_{\bf k}^2
\, ,
\label{wronskianspin}
\ear
and that ${\cal N}_\bk(t)$ can be written in terms of the solutions of the mode equation as
\bear
{\cal N}_\bk(t) &=& \frac{1}{2\omega_\bk (\omega_\bk - \Pi_{\bf k})}
\biggl\lbrack
\omega_\bk^2 |\psi_{\bf k}^{(\pm)}|^2 + |\dot \psi_\bk^{(\pm)}|^2 \mp i\omega_\bk
(\dot\psi_{\bf k}^{(\pm)} \psi_{\bf k}^{(\pm)\ast} - \psi_{\bf k}^{(\pm)}\dot\psi_{\bf k}^{(\pm)\ast})
\biggr\rbrack
\label{Nspin}
\ear
(as a consequence of parity invariance, both spin projections give the same density of created pairs).
Combining the last two equations, we can also write
\bear
{\cal N}_\bk(t) &=& \frac{\mu_{\bf k}^2}{2\Pi_{\bf k} (\omega_\bk - \Pi_{\bf k})}
-
\frac{\omega_\bk^2 |\psi_\bk^{(\pm)}|^2 + |\dot \psi_\bk^{(\pm)}|^2}
{2\omega_\bk\Pi_{\bf k}} \, .
\ear
Defining $G_\bk \equiv \vert\psi_\bk^{(+)}\vert^2$ like in the scalar case
(we choose the upper spin component for definiteness)
and combining \eqref{meqspin} and \eqref{wronskianspin}, one arrives
at the following fermionic generalization of the Gelfand-Dikii equation \eqref{geldikeq}:
\begin{equation}
\dddot{G_\bk}-\frac{qE}{\Pi_{\bf k}}\ddot{G_\bk}+4\omega_\bk^2\dot{G_\bk}
+\Bigl(4\dot\omega_{\bf k}\omega_{\bf k}
-4\frac{qE\omega_\bk^2}{\Pi_{\bf k}}\Bigr)G_\bk=-2\mu_{\bf k}^2\frac{qE}{\Pi_{\bf k}}
\, .
\label{geldikeqspin}
\end{equation}
Finally, differentiating the mode equation one can eliminate $ |\dot \psi_\bk^{(+)}|^2$ in the same
way as in \eqref{elphidot}, and rewrite ${\cal N}_\bk$ in terms of $G_\bk$:
\bear
{\cal N} _\bk= \frac{1}{2} + \frac{\omega_\bk}{\Pi_{\bf k}}\Bigl(\frac{1}{2}-G_\bk\Bigr) - \frac{\ddot G_\bk}{4\omega_\bk \Pi_{\bf k}}
\, .
\label{NspinG}
\ear

\section{Solitonic fields in scalar QED}
\label{sec:Solitonic fields in scalar QED}

\subsection{Definition and properties}

Next, let us return to the scalar case and study the solitonic fields \eqref{Anscal}. For $\bf k = \tilde {\bf k}$
the frequencies are given by
\begin{eqnarray}
\omega_{(p)}^2 (t) = \omega_0^2 + \frac{p(p+1) \omega_0^2}{\cosh^2 (\omega_0 t)} \, . \label{1 sol freq}
\end{eqnarray}
Thus after a rescaling $\omega_0t\to t$ the corresponding mode equation \eqref{meqscal}
can be rewritten as the Schr\"odinger equation
\begin{equation}\label{sch}
\ddot\phi+(E-V(t))\phi=0 \, ,
\end{equation}
with $E=1$ (not to be confused with the electric field strength) and the P\"oschl-Teller potential
\begin{equation}\label{V}
V(t)=-\frac{p(p+1)}{\cosh^2 t} \, .
\end{equation}
By the above-mentioned connection between the pair-creation and over-the-barrier scattering problems,
particle production is absent for integer values of $p$ where the P\"oschl-Teller potential becomes
reflectionless.
The solution of the mode equation is reviewed in the appendix, eq. \eqref{solsol}. The corresponding
solution of the Gelfand-Dikii equation \eqref{geldikeq} is
\bear
G_{(p)}
&=&
|\phi_{(p)}|^2
=
\frac{1}{2\omega_0}\,(1+\e^{2\omega_0 t})^{2p+2} |\2F1 (p+1-i,p+1,1-i;-\e^{2\omega_0 t})|^2
\, .
\label{exactsol}
\ear
Remarkably, for natural $p$ the Gelfand-Dikii equation
can also be solved globally in terms of hyperbolic functions \cite{Kim:2011jw},
using the ansatz
\begin{eqnarray}
G_{(p)} = \frac{1}{\omega_0} \sum_{n = 0}^\infty C_n \frac{1}{\cosh^{2n} (\omega_0 t)}, \qquad C_0 = \frac{1}{2} \, , \label{ser sol}
\end{eqnarray}
where the first term is determined by the initial condition \eqref{Ginitial}. We find the recursive relation
\begin{eqnarray}
C_n = \frac{(2n-1)}{(2n)} \,\frac{n(n+1) - p(p+1)}{(n^2+1)} C_{n-1} \, . \label{rec rel}
\end{eqnarray}
For natural $p$, all the coefficients $C_n$ vanish for $n \geq p+1$.
Let us give these solutions explicitly for the first three cases $p=1,2,3$:
\vspace{10pt}
\begin{eqnarray}
G_{(1)}(t) &=& \frac{1}{2 \omega_0} \Big(1- \frac{1}{2} \frac{1}{\cosh^2 (\omega_0 t)}\Big) \, ,\nonumber\\
G_{(2)}(t) &=& \frac{1}{2 \omega_0}\Big(1 - \frac{3}{2} \frac{1}{ \cosh^2 (\omega_0 t)} + \frac{9}{10} \frac{1}{ \cosh^4 (\omega_0 t)}\Big)\,, \nonumber\\
G_{(3)}(t) &=& \frac{1}{2 \omega_0}\Big(1 -\frac{3}{ \cosh^2 (\omega_0 t)} + \frac{9}{2} \frac{1}{ \cosh^4 (\omega_0 t)}-\frac{9}{4} \frac{1}{ \cosh^6 (\omega_0 t)}\Big) \, .
\label{C123}
\end{eqnarray}
Note that (\ref{exactsol}) directly gives (\ref{C123}) as expected.
Since these polynomials depend on $t$ only through $\cosh(\omega_0 t)$, plugging them into \eqref{NtoG} obviously yields
${\cal N}_{\tilde \bk}(t)$ that are time-inversion symmetric, to that their vanishing for $t\to -\infty$ implies the same for $t\to\infty$.

However, this raises the question of why the same is no longer true for non-integer $p$, where pair-creation is known to occur
from the exact solution to the Klein-Gordon equation \cite{Kim:2011jw}.
In that case the recursive relation (\ref{rec rel}) does not terminate, and asymptotically leads to
\bear
C_n /C_{n-1}\simeq 1 - \frac{3}{2n} + O\Bigl(\frac{1}{n^2}\Bigr)\,,
\label{asymptcoeff}
\ear
where the two leading terms shown are independent of $p$. From the leading term we conclude
that $G_{(p)}$ as a power series in $z=1/\cosh^2(\omega_0t)$ has unit convergence radius.
From the subleading one it follows that, independently of the value of $p$ (as long as it is non-natural), the coefficients fall off like
$C_n \sim n^{-\frac{3}{2}}$, so that the series still converges for $z=1$, and thus for all $t$.

\begin{figure}[!htbp]
\vspace{-250pt}
        \centering
      \includegraphics[height=\textheight]{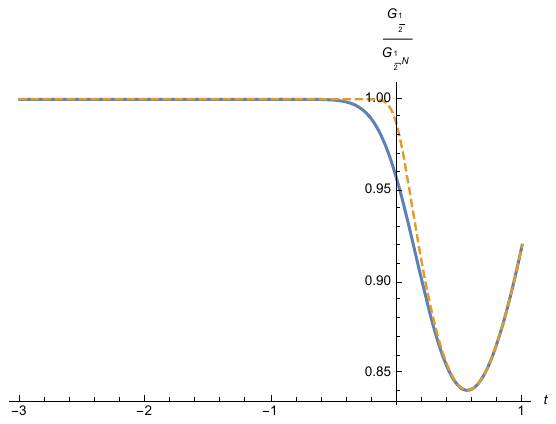}
       \vspace{-270pt}
     \caption{Plot of the ratio $G_{(p)}(t)/G_{(p),N}(t)$, where $G_{(p)}$ is the exact solution \eqref{exactsol} and  $G_{(p),N}$
    the power series \eqref{ser sol} truncated to run only up to $n=N$. The continuous line is for $N=10$ and the dashed line for $N=100$.
     }
      \label{fig3}
\end{figure}

Nevertheless, the fact that at $t=0$ it touches the boundary of the region of convergence is sufficient to invalidate the uniqueness of the
analytic continuation beyond this time, as is illustrated in Fig. \ref{fig3}. Here we plot, for $p=\half$, the ratio between
 the exact solution \eqref{exactsol} and the power series \eqref{ser sol} truncated to run only up to $n=N$.
The result clearly suggests that, if it was possible to remove the cutoff $N$, the ratio would be equal to unity up to $t=0$ but not beyond.

\vspace{100pt}

\subsection{Numerical solution of the mode equation}

\begin{figure}[!htbp]
    \centering
    \begin{minipage}{.5\textwidth}
        \centering
        \includegraphics[width=0.8\linewidth, height=0.15\textheight]{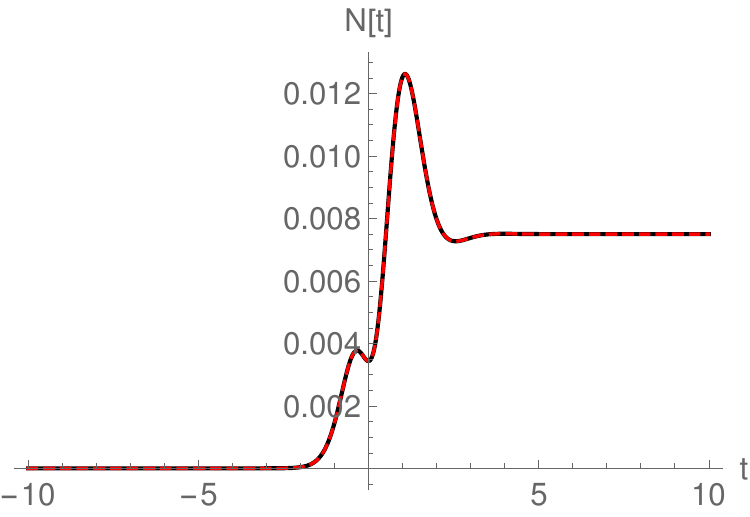}
    \end{minipage}%
    \begin{minipage}{0.5\textwidth}
        \centering
        \includegraphics[width=0.8\linewidth, height=0.15\textheight]{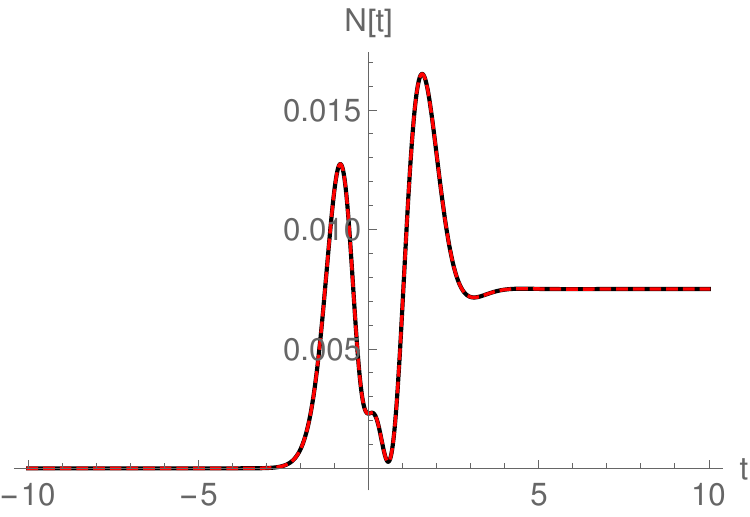}
    \end{minipage}
    \caption{Comparison between the exact (black line) and numerical (dashed red-line) calculations of ${\cal N}_{\tilde\bk}(t)$
        for the solitonic background. The left panel is for $p=\half$ and the right panel for $p=\frac{3}{2}$. We set $\omega_0=1.1$ and $k_\parallel=\tilde k_\parallel = 1$. Clearly at $t\rightarrow\infty$
    there is pair production. }
    \label{fig9}
\end{figure}

\begin{figure}[!htbp]
    \centering
    \begin{minipage}{.5\textwidth}
        \centering
        \includegraphics[width=0.8\linewidth, height=0.15\textheight]{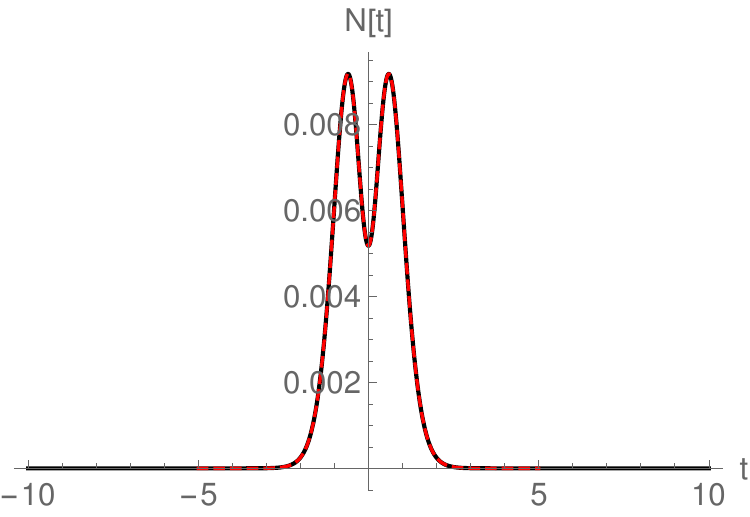}
    \end{minipage}%
    \begin{minipage}{0.5\textwidth}
        \centering
        \includegraphics[width=0.8\linewidth, height=0.15\textheight]{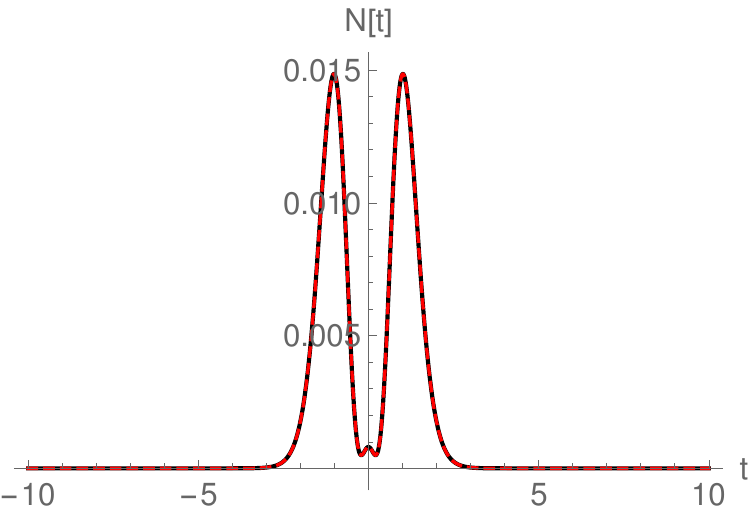}
    \end{minipage}
      \caption{Comparison between the exact (black line) and numerical (dashed red-line) calculations of ${\cal N}_{\tilde\bk}(t)$
        for the solitonic background. The left panel is for $p=1$ and the right panel for $p=2$. We set $\omega_0=1.1$ and $k_\parallel=\tilde k_\parallel = 1$.
        The plots clearly show the time-inversion symmetry and the absence of pair production. }
    \label{fig10}
\end{figure}

As a warm-up for the fermionic case, let us now compare ${\cal N}_{\tilde\bk}(t)$ as obtained from
the above exact solutions of the mode equation used in \eqref{NtoG} vs a numerical calculation using the Vlasov equation \eqref{inout}.
In Figs. \ref{fig9}, \ref{fig10} we show the results for some special values of
the P\"oschl-Teller parameter $p$, integer as well as non-integer ones.
For all the cases the match is excellent. 


\subsection{Solitonic pair creation for general momenta}

\begin{figure}[!htbp]
\center{\includegraphics[width=9.0cm]{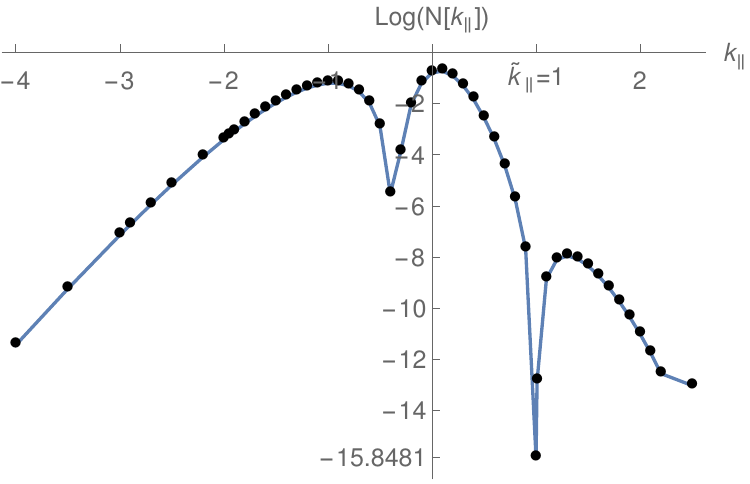}}
\caption{Scalar pair-creation spectrum for the solitonic field  \eqref{Anscal}
with $p=1$, reference momentum $\tilde{k}_\parallel=1$, and $\mu_{\bf k}^2=\mu_{\bf \tilde k}^2=0.21$, i.e. $\omega_0^2=k_\parallel^2+0.21$.
Here the lowest point shows the vanishing of the density of created pairs at $k_\parallel=\tilde k_\parallel=1$.}
\label{fig5}
\end{figure}

It must be emphasized that, even when the solitonic condition of integer $p$ is met, this implies
a vanishing density of created pairs only for $\bf k = \tilde {\bf k}$. For momenta different
from the reference momentum $\tilde{\bf k}$ we do expect pair creation, particularly since there exist general arguments
that the total density of created pairs for a purely time-dependent field can never vanish \cite{63}.
In Fig. \ref{fig5} we show the result of a numerical evaluation of the density of created pairs
as a function of $k_{\parallel}$ for the field \eqref{Anscal} with $p=1$, showing that the density of created pairs 
vanishes at the chosen reference momentum $\tilde k_\parallel$, but has a quite complicated structure otherwise.



\section{Solitonic fields in spinor QED: the original solitons}
\label{sec:Solitonic fields in spinor QED: the original solitons}

At this stage, it is natural to ask whether the concept of an electric field that can be tuned not to pair create
at some given reference momentum can be generalized to the spinor QED case.
Using the solitonic fields \eqref{Anscal} in the fermionic mode equation \eqref{meqspin} leads
to a corresponding Schr\"odinger equation that, even for $\bf k = \tilde {\bf k}$,
seems analytically intractable.

Nonetheless, as reported in \cite{lphys}, a search based on the numerical solution of the fermionic evolution
equation \eqref{inoutspin} led to certain ``magic values'' of the index $p$,
\bear
p = \half \Bigl(-1 \pm \sqrt{m^2+m+1}\Bigr), \quad m=2,4,6,\ldots\,,
\label{magic}
\ear
where ${\cal N}_{\tilde\bk}(t)$ returns to zero for $t\to\infty$, as far as one can tell
from the numerical analysis. However, further study showed that these values are not universal, but depend
on the reference momentum. We find that, heuristically, the magic values can be parametrized as
\bear
p(p+1)=n(n+x(k_\parallel))\, , \quad n=0,1,2,3,\cdots\,,
\label{parametrize}
\ear
where $x(k_\parallel)$ is a monotonically increasing function with
\bear
x(0) = 0 \, ,
\label{x0}
\ear
and it is always understood that $\bf k = \tilde {\bf k}$.
Some numerically determined values of $x(k_\parallel)$ are given in Table \ref{table}.

\bigskip
\begin{table}
\caption{Numerically experimented $x(k_\parallel)$ for different values of $k_\parallel$ and fixed $\mu_{\bf k}=1$.}
\begin{tabularx}{1\textwidth}
{
  | >{\raggedright\arraybackslash}X
  | >{\centering\arraybackslash}X
  | >{\raggedleft\arraybackslash}X
   | >{\raggedleft\arraybackslash}X
    | >{\raggedleft\arraybackslash}X
     | >{\raggedleft\arraybackslash}X
      | >{\raggedleft\arraybackslash}X
       | >{\raggedleft\arraybackslash}X
        | >{\raggedleft\arraybackslash}X
         | >{\raggedleft\arraybackslash}X
          | >{\raggedleft\arraybackslash}X
          | >{\raggedleft\arraybackslash}X
            | >{\raggedleft\arraybackslash}X
              | >{\raggedleft\arraybackslash}X
                | >{\raggedleft\arraybackslash}X
                | >{\raggedleft\arraybackslash}X
                  | >{\raggedleft\arraybackslash}X
                    | >{\raggedleft\arraybackslash}X | }
 \hline
 $k_\parallel$ & 0 & 0.1 &0.2&0.3&0.4&0.5&0.6&0.7&0.8&0.9&1&2&3&4&5&6&7\\
 \hline
$ \hspace{-0.1cm}x(k_\parallel)$  & 0 &$10^{-3}$&0.02& 0.1&0.18& 0.25&0.32&0.35&0.36&0.4&0.44&0.54&0.56&0.58&0.58&0.58&0.59 \\
\hline
\end{tabularx}

\label{table}
\end{table}

The relation between \eqref{parametrize} and the original \eqref{magic} is the following. Before it was realized that the magic values depend on $k_{\parallel}$,
the numerical calculations leading to \eqref{magic} were all performed using the fixed value $k_{\parallel}=1$. According to \eqref{parametrize} and Table \ref{table}
this leads to $p(p+1) = n(n+ 0.44)$ which is numerically close to $p(p+1) = n(n+ \frac{1}{2})$ as obtained from \eqref{magic} for $m=2n$.

\begin{figure}[!htbp]
\vspace{32pt}
\center{\includegraphics[width=7cm]{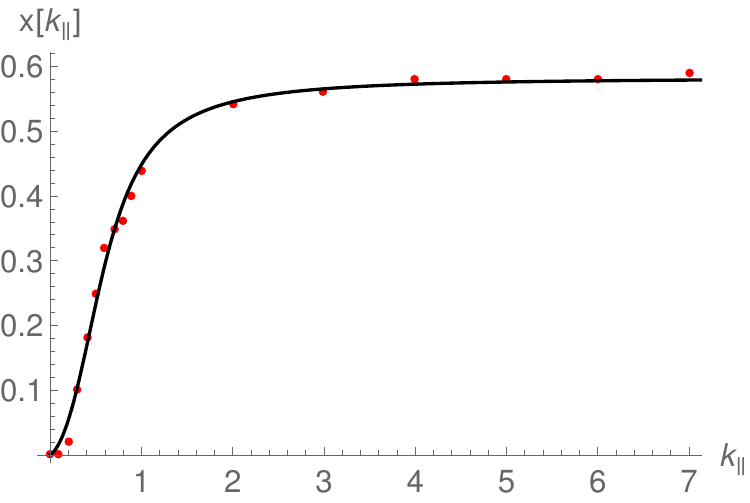}}
\caption{Fit of the values of $x(k_\parallel)$ given in Table \ref{table} (red points) against Eq. (\ref{fit}) (solid curve). }
\label{fig-x-k}
\end{figure}

In Fig. \ref{fig-x-k} we plot the values for $x(k_\parallel)$ given in Table \ref{table}, and we observe that they can be
nicely fitted using the ansatz
\bear
x(k_\parallel) = a\arctan(bk_\parallel+ck_\parallel^2) \, ,
\label{fit}
\ear
with
$
a=0.370315\, ,b=0.2537\, , c=2.385 \, .
$

In Figs. 6 and 7 we show some sample plots, using \eqref{inoutspin}, for parameter values where no pairs are created, as far as one can tell
from a numerical analysis.

\vspace{30pt}

\begin{figure}[h]
    \centering
    \begin{minipage}{.5\textwidth}
        \centering
        \includegraphics[width=0.9\linewidth, height=0.15\textheight]{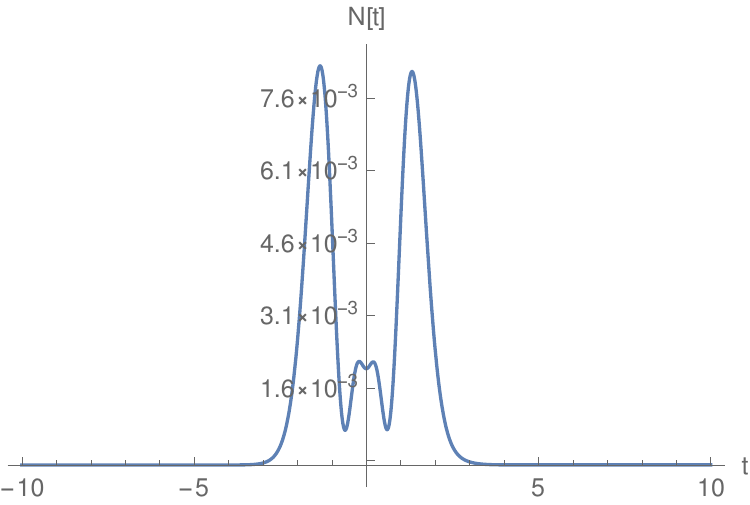}
    \end{minipage}%
    \begin{minipage}{0.5\textwidth}
        \centering
        \includegraphics[width=0.9\linewidth, height=0.15\textheight]{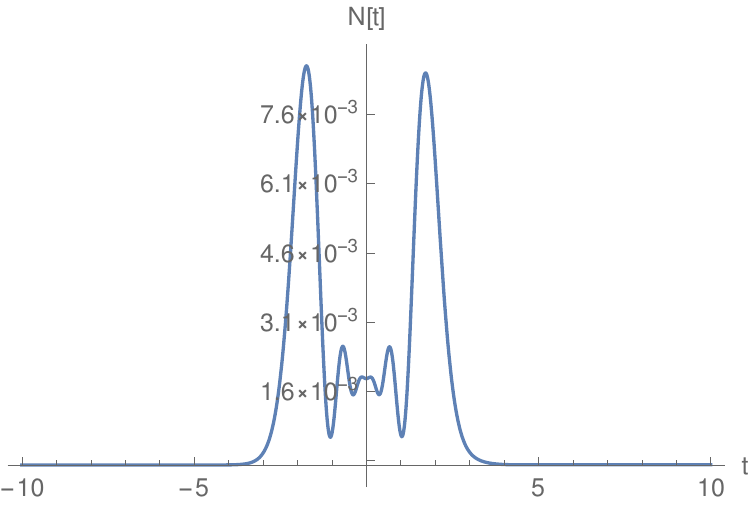}
    \end{minipage}
    \caption{Plots for $k_\parallel=0.7$, $\mu_{\bf k}=1$, left for $n=3$ and right for $n=5$, based on Eq. \eqref{parametrize}.}

\vspace{30pt}

    \begin{minipage}{.5\textwidth}
        \centering
        \includegraphics[width=0.9\linewidth, height=0.15\textheight]{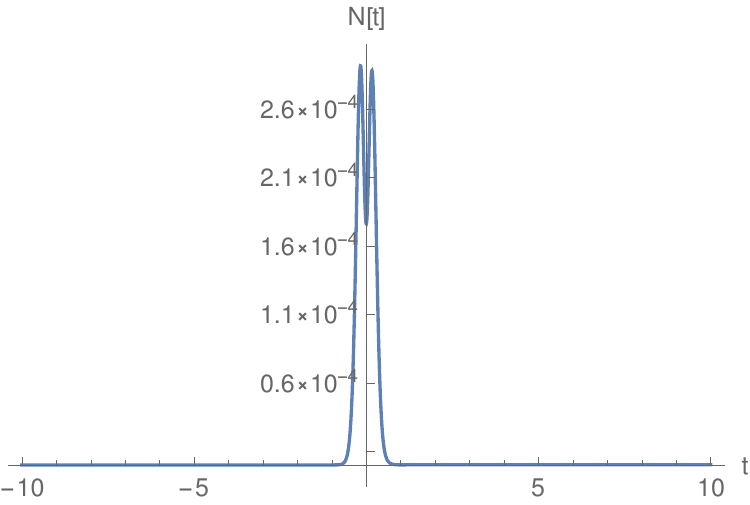}
    \end{minipage}%
    \begin{minipage}{0.5\textwidth}
        \centering
        \includegraphics[width=0.9\linewidth, height=0.15\textheight]{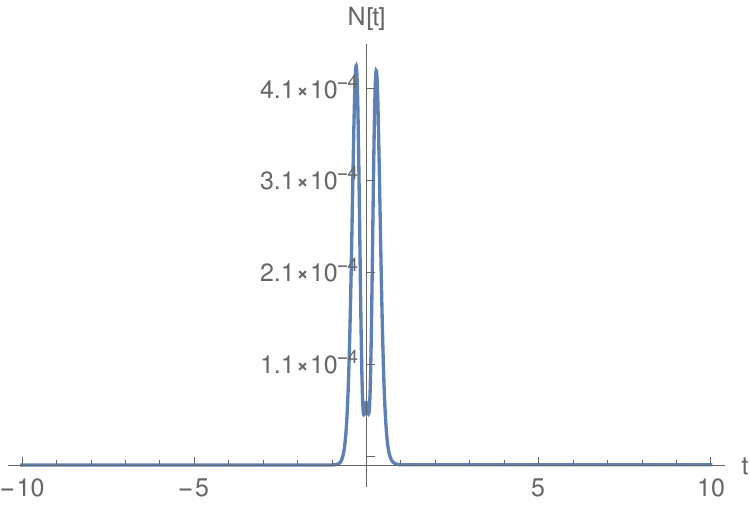}
    \end{minipage}
      \caption{Plots for $k_\parallel=4$, $\mu_{\bf k}=1$, left for $n=1$ and right for $n=2$, based on Eq. \eqref{parametrize}.}    
\end{figure}

\begin{center}

\begin{figure}[!htbp]
\vspace{12pt}
\center{\includegraphics[width=7.0cm]{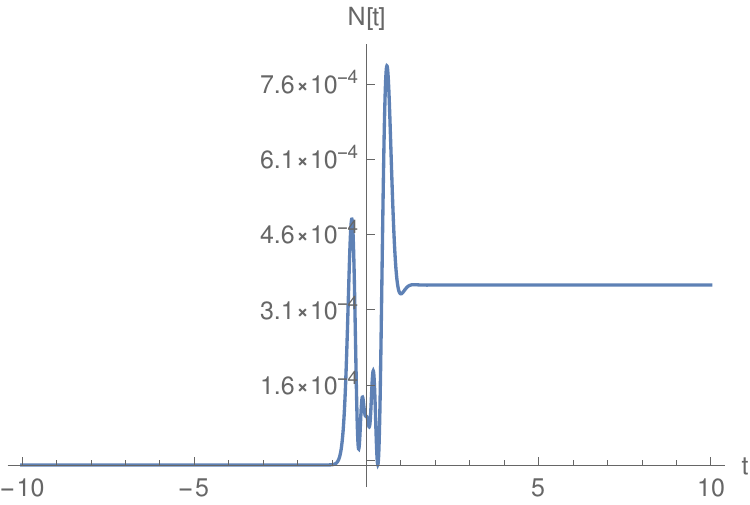}}
\caption{Plot for the generic parameters $k_\parallel=4$, $\mu_{\bf k}=1$, and $n=3$, showing pair creation.}
\label{fig-solk}
\end{figure}

\end{center}

This should be contrasted with the plot in Fig. \ref{fig-solk} for generic pair-creating parameters.

\section{Solitonic fields in spinor QED: the alternative solitons}
\label{sec:Solitonic fields in spinor QED: the alternative solitons}
Finally, let us introduce here a different definition of the solitonic fields that in the spinor case we have found more tractable analytically than the original one.
The new definition is
\begin{equation}\label{eA2}
qA^{\rm new}_{\parallel(p)}(t)\equiv \tilde{k}_\parallel-\frac{\sqrt{p(p+1)}\mu_{\tilde \bk}}{\cosh(\mu_{\tilde \bk} t)} \, .
\end{equation}
The corresponding scalar mode equation \eqref{meqscal} leads, for $\bf k = \tilde {\bf k}$
and this time using the rescaling $\mu_{\tilde\bk} t \to t$, to the same Schr\"odinger equation \eqref{sch}
that we had obtained with the original definition \eqref{Anscal},
and thus to the same pair non-creation properties. However, whenever there is pair creation the rates will in general be different, since the time rescaling does not leave the initial
conditions invariant.

In spinor QED the two definitions are (except for $ k_{\parallel} = \tilde k_{\parallel} =0$, where also $\omega_0=\mu_{\tilde\bk}$)
inequivalent, and the new one has an advantage in that it still allows one to solve the mode equation exactly.
If we substitute (\ref{eA2}) into \eqref{meqspin} (for $\bf k = \tilde {\bf k}$ and choosing the lower sign),
and rescale $\mu_{\tilde\bk} t \to t$, we again get the Schr\"odinger equation (\ref{sch}), with $E=1$ and now with
the potential
\begin{equation}\label{Vf}
V(t)=-\frac{p(p+1)}{\cosh^2t}-i\frac{\sqrt{p(p+1)}\tanh t}{\cosh t} \, ,
\end{equation}
which was studied in \cite{khare_jpa1988}.
According to (\ref{bcferm}) the pair creation probability reads
\begin{equation}
{\rm lim}_{t\to\infty}
 {\cal N}_{\tilde\bk}(t)
=\frac{\sin^2(\pi\sqrt{p(p+1)})}{\cosh^2\pi } \, .
\end{equation}
Therefore the condition for pair non-creation for the alternative fermionic solitons becomes
\begin{equation}
p(p+1)=n^2\Rightarrow p=\half(-1\pm\sqrt{1+4n^2})\, ,~~~n=1,2,3,\cdots\,,
\end{equation}
independently of the choice of $\bf k = \tilde {\bf k}$.
Since for $\tilde k_{\parallel} = k_{\parallel} =0$ there is no difference between the
original and the alternative solitons, this also provides an analytical confirmation of \eqref{x0}.

\begin{figure}[!htbp]
    \centering
    \begin{minipage}{.5\textwidth}
        \centering
        \includegraphics[width=0.8\linewidth, height=0.15\textheight]{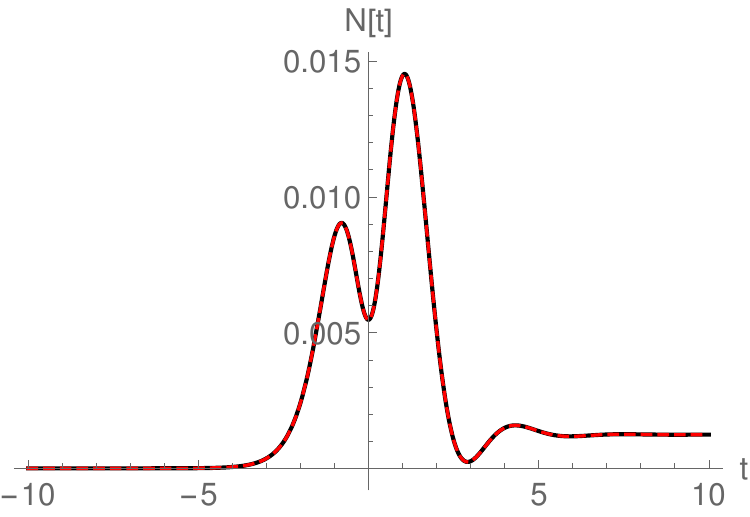}
    \end{minipage}%
    \begin{minipage}{0.5\textwidth}
        \centering
        \includegraphics[width=0.8\linewidth, height=0.15\textheight]{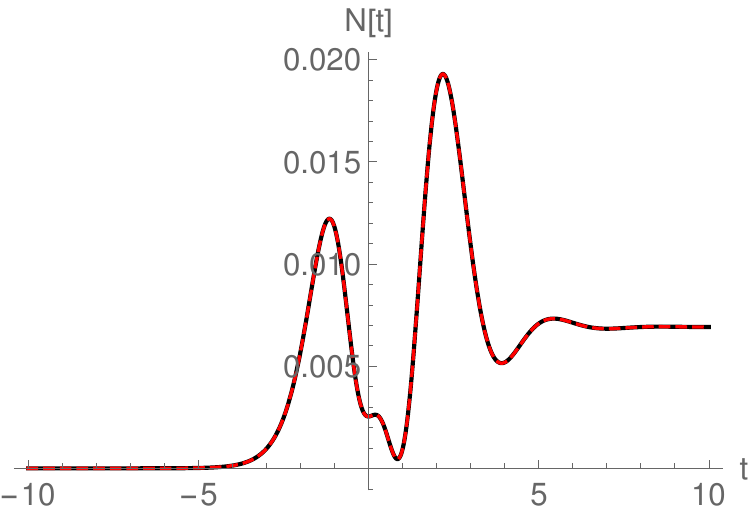}
    \end{minipage}
    \caption{Comparison between the exact (black line) and numerical (dashed red-line) calculations of ${\cal N}_{\tilde\bk}(t)$
        for the alternative fermionic soliton background. Here we set $k_\parallel=0, \omega_0=\mu_{\bf k}=1$. The left panel is
        for $p=\half$, the right panel for $p=1$, both non-magic values showing pair production at $t\rightarrow\infty$.}
        \label{figFpairs}
\end{figure}

\begin{figure}[h]
    \centering
    \begin{minipage}{.5\textwidth}
        \centering
        \includegraphics[width=0.8\linewidth, height=0.15\textheight]{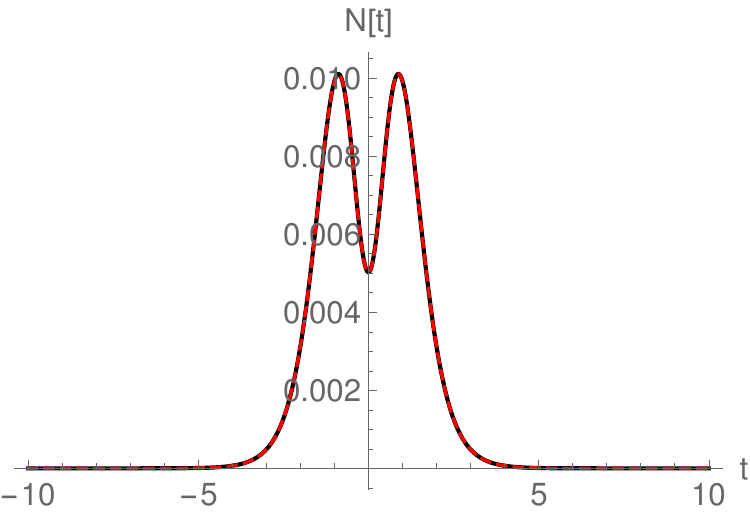}
    \end{minipage}%
    \begin{minipage}{0.5\textwidth}
        \centering
        \includegraphics[width=0.8\linewidth, height=0.15\textheight]{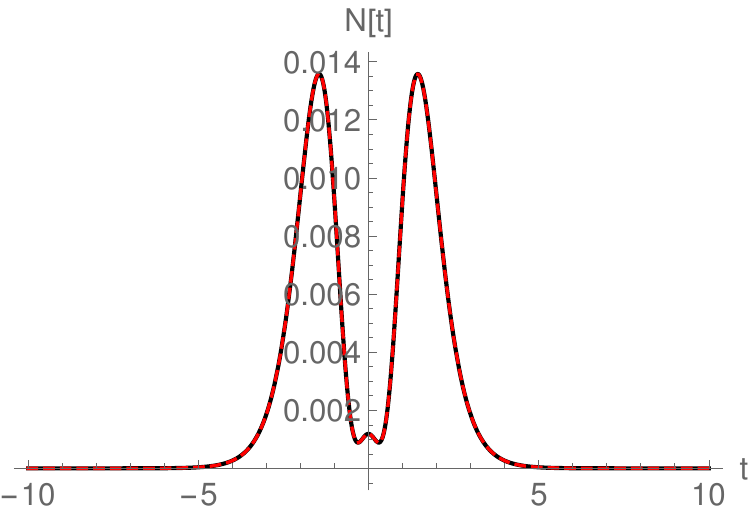}
    \end{minipage}
    \caption{Comparison between the exact (black line) and numerical (dashed red-line) calculations of ${\cal N}_{\tilde\bk}(t)$
        for the alternative fermionic soliton background. Here we set $k_\parallel=0, \omega_0=\mu_{\bf k}=1$. The left panel is for $p=\half(\sqrt{5}-1)$, the right panel for $p=\half(\sqrt{17}-1)$, which are magic values and give perfectly symmetric plots and no pairs at $t\rightarrow\infty$.}
           \label{figFnopairs}
\end{figure}

In Figs. \ref{figFpairs} and \ref{figFnopairs} we once more compare results obtained for ${\cal N}_{\tilde {\bf k}}(t)$ using the
exact solution \eqref{psinewsol} in \eqref{NspinG} vs the fermionic Vlasov equation \eqref{inoutspin}, finding
excellent agreement for two pair-creating cases (Fig. \ref{figFpairs}) and two non-creating ones (Fig. \ref{figFnopairs}).

\section{Solitonic pair non-creation as a Stokes phenomenon}
\label{sec:Solitonic pair non-creation as a Stokes phenomenon}

\begin{figure}
\includegraphics[scale=0.16]{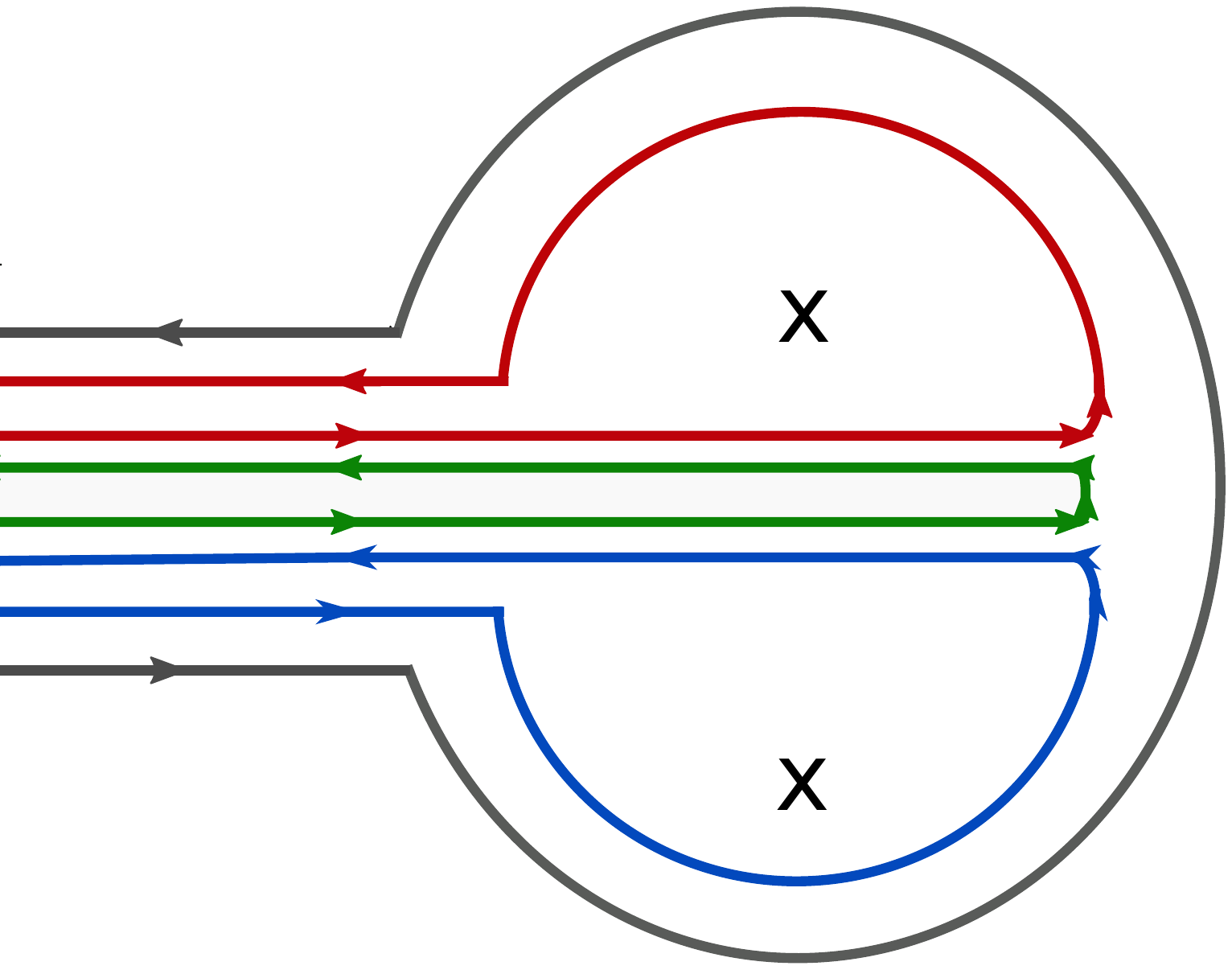}
\caption{Four contours of winding number one enclosing or excluding simple poles at $z=\pm i$. The contour in black encloses both poles, the blue and red include only one of the poles, and the green contour excludes both poles.}
\label{counter}
\end{figure}

Another advantage of the alternative definition of the solitonic fields is that it allows us to understand the
physical origin of the pair no-creation as a quantum interference phenomenon,
using the phase-integral formulation. In the phase-integral formulation~\cite{Kim:2007pm}, Fourier modes for a quantum field equation in a one-dimensional background field are extended to a complex plane of time or space, and the leading behavior of pair production for each mode is determined by simple poles and their residues. The spin-diagonal fermionic mode equations~\eqref{meqspin}  take the form
\begin{eqnarray}
\ddot{\psi}^{(\pm)} (t) + \omega^{(\pm) 2} (t)  \psi^{(\pm)} (t) = 0
~~~,~~~ \omega^{(\pm) 2} (t) = \frac{b^2}{\cosh^2t} \pm i \frac{b \tanh t}{\cosh t} \,,
\end{eqnarray}
where $b = \sqrt{p(p+1)}$.
The leading terms for the mean number for scalar pair production~\cite{Kim:2013cka} can be extended to spinor pair production,
\begin{eqnarray}
{\cal N}^{(\pm)} \simeq \Bigl{|} \sum_{J} \exp [- i \oint_{C^{(1)}_{J (-\infty)} } \omega^{(\pm)} (t) dt ] \Bigr{|} \,,
\label{cont-mean}
\end{eqnarray}
where the contour $C^{(1)}_{J (-\infty)}$ either enclose finite simple poles, or exclude them entirely while starting at the past infinity ($ t = - \infty $) (the $\simeq$ sing in (\ref{cont-mean}) refers to the fact that we restrict the sum to contours of winding number 1). It is important to note that there might be another simple pole at $t = \infty$, which is excluded by contours but contributes a common factor.

Under a conformal transformation $z = e^{t}$, the contour integral becomes
\begin{eqnarray}
\oint_{C^{(1)}_{J (-\infty)} } \frac{dz}{z} \frac{\sqrt{(z^2+1)^2 + 4 b^2 z^2 \pm 2 i b z (z^2 -1)}}{z^2 +1} \,.
\end{eqnarray}
The integrand can be made an analytic function by properly choosing branch cuts. First, there is a simple pole at $z = \infty$, whose residue gives an overall factor $e^{-2 \pi}$. Second, two finite simple poles are located at $z = i$ and $z = -i$, whose residues are $i \sqrt{b^2 + b} = i (b+ 1/2 + \cdots) $ and $-i \sqrt{b^2 - b} = - i (b - 1/2 + \cdots) $, respectively, for asymptotically large $b$ expansion. Then, the mean number~\eqref{cont-mean} is given by
\begin{eqnarray}
{\cal N}^{(\pm)} \simeq e^{- 2 \pi} \Bigl( 1+ e^{2 \pi i (b + 1/2) - 2 \pi i (b - 1/2)} + e^{2 \pi i (b + 1/2)} + e^{-2 \pi i (b - 1/2)} \Bigr) \,.
\end{eqnarray}
Here, the first term in the parenthesis comes from a contour excluding both $z = \pm i$ while the second term comes from a contour enclosing both poles. The third and fourth terms come from contours enclosing one pole each, as shown in Fig. \ref{counter}. Simple arithmetic leads to
\begin{eqnarray}
{\cal N}^{(\pm)} \simeq e^{- 2 \pi} (2 \sin \pi b)^2 \,,
\end{eqnarray}
which is the leading term of (\ref{bcferm}). A passing remark is that for a scalar field, the finite residues are simply $\pm 2 \pi i b$, which give ${\cal N} \simeq e^{- 2 \pi} (2 \cos \pi b)^2$.

The solitonic nature of pair production can be understood as the Stokes phenomenon~\cite{Kim:2013cka}. For instance, a massive scalar field in the global coordinates of de Sitter (dS) space has two simple poles: the north pole and the south pole. Pairs of particles produced from each pole interfere destructively or constructively, which depends on the dimension of spacetime. No particles are produced in any odd-dimensional dS space. On the contrary, a massive scalar field in the planar coordinates of the dS space has only one pole, the north or south pole, which always produces particle pairs regardless of dimensions. This Stokes phenomenon has been confirmed by field theory~\cite{Mottola1985,Bousso:2001mw}. Recently it was shown in Ref.~\cite{Jiang:2020evx} that fermion production exhibits the same Stokes phenomenon in the global dS spaces.

A physical interpretation of the solitonic nature of particle production in time-dependent electric fields is the existence of a reflectionless potential, which means that a positive frequency solution from the asymptotic past scatters over the potential barrier without reflection to another positive frequency solution in the asymptotic future. In the bosonic theory, the reflectionless potentials are closely related to supersymmetric quantum mechanics and give the instantaneous snapshot of a family of solitons~\cite{Schonfeld1980,Kwong1989}. Physically, when the frequency has more than two pairs of turning points in the complex plane of time, superposing instanton paths connecting pairs of turning points leads to a sinusoidal behavior of particle production and thus Stokes phenomena for certain parameters~\cite{Dumlu:2010ua}. Here we illustrated the Stokes phenomenon in the phase-integral formulation.
%

\section{Conclusions}
\label{sec:Conclusions}

We have extended to the spinor QED case an approach to Schwinger pair creation for purely time-dependent fields that
is based on the Gelfand-Dikii equation rather than the mode equation or the Vlasov equation.
Replacing the mode equation, an equation for $\psi^{(\pm)}_{\bf k}(t)$, by the Gelfand-Dikii equation, which is an equation for $\vert{\psi^{(\pm)}_{\bf k}}(t)\vert^2$,
is suggested by the fact that the phase of $\psi^{(\pm)}_{\bf k}(t)$ is known to be redundant in the pair-creation context.
As in the scalar QED case, the time evolution is then described by a real third-order linear differential equation, as compared
to a second-order complex equation or an integro-differential equation. We hope to explore the usefulness of this equation for a direct
numerical evaluation of density of created pairs in the future.
Here, our focus was on deriving this ``fermionic Gelfand-Dikii equation'', and searching for a fermionic generalization
of the solitonic fields that in previous work on scalar QED had been discovered through the relation between the Gelfand-Dikii equation and the KdV equation.

For the original solitonic fields, we have provided ample numerical evidence that they become pair non-creating
for fermions by changing the parameter $p$ from integer to certain other ``magic'' values. However, these parameters turn out to depend on the choice of the reference momentum,
and we were able to determine them only in a heuristic and approximate manner.
This has led us to introduce an alternative family of solitonic fields that allow for an exact analytical
treatment, resulting in the simple and universal criterion $p(p+1)=n^2$ for the absence of  pair creation.
Thus for any given reference momentum we are now in a position to construct electric fields that will create scalar particles but not spinor particles,
and also the other way round. 
Apart from their intrinsic physical interest, our soliton solutions may also become useful by providing a benchmark for numerical methods.

We leave it to future work to search for some generalization of the KdV equation that hopefully would allow one to link our fermionic
generalizations of the Gelfand-Dikii equation and of the solitonic fields in a way similar to what had emerged in \cite{Kim:2011jw}
for the scalar case.

Other issues in the fermionic case that call for further investigation are (i) whether and how the solution of the scalar Gelfand-Dikii equation of section V 
as a series in powers of $1/{\rm cosh}(\omega_0 t)$ can be generalied to the fermionic equation \eqref{geldikeqspin} (ii) for the original solitons, to find
a physically more satisfactory derivation of the ``magic numbers'' and (iii) to explain the behavior of $x(k_{\parallel})$ found heuristically in \eqref{fit}, or at
least its apparent convergence for large $k_{\parallel}$.

\acknowledgments
N.~A. would like to thank C. Kohlf\"urst and S. Lang for valuable discussion especially on numerical studies as well as R. Sch\"utzhold for discussion and support.
C.~S. would like to appreciate the warm hospitality at and support by National Research Foundation (NRF) of Republic of Korea funded by the Ministry of Education (2019R1I1A3A01063183) through Kunsan National University, where this work was initiated, and revised, 
and S.~P.~K. would like to appreciate the warm hospitality at Instituto de F\'{\i}sica y Matem\'aticas,
Universidad Michoacana de San Nicol\'as de Hidalgo, and the Helmholtz-Zentrum Dresden-Rossendorf. 
A.~M.~F. thanks the Centro Internacional de Ciencias A.C., UNAM-UAEM, for hospitality.
S.~P.~K.  was supported by IBS (Institute for Basic Science) of Republic of Korea under IBS-R012-D1.
E.~G.~G was supported by the project ADONIS (Advanced research using high intensity laser produced photons and particles) CZ.02.1.01/0.0/0.0/16\_019/0000789 from European Regional Development Fund.
A.~M.~F. was supported by the Russian Foundation for Basic Research (Grant No. 20-52-12046).
Finally, our sincere thanks to the anonymous referee for a number of useful suggestions.

\appendix

\section{Solving the mode equations}

In this appendix we solve the scalar and spinor QED mode equations
\eqref{meqscal} and \eqref{meqspin} for solitonic fields, including also the benchmark Sauter-field case
for easy reference. The subscript $\bk$ is omitted throughout.

\subsection{Solution of the mode equations for the Sauter field}
\label{app-solsauter}

As a warm-up, let us solve the scalar and spinor mode equations
\eqref{meqscal} and \eqref{meqspin} for the benchmark case of the time-like Sauter field, defined by
\bear
E(t)=-\dot{A} = -E_0 \,{\rm sech}^2(t/\tau) \, ,
\label{defEsauter}
\ear
which we can realize by
\bear
A^{\mu} = (0,0,0,E_0 \tau (1+\tanh (t/\tau) ))
\, .
\label{defAsauter}
\ear
Thus we have
\bear
\omega^2(t) = \Bigl\lbrack k_{\parallel} - qE_0 \tau  (1+\tanh (t/\tau) )\Bigr\rbrack^2 + \mu^2
\ear
and
\bear
\omega_i&=&\omega_0 \equiv \sqrt{{\bf k}^2+m^2} \, ,\\
\omega_f&=& \sqrt{(k_{\parallel}-2qE_0\tau)^2+\mu^2} \, .
\ear
We can combine the scalar and spinor QED cases as
\begin{eqnarray}
\ddot{\psi}^{\sigma} (t) + [\omega^2 (t) - 2 \sigma iq \dot A_{\parallel}(t)]  \psi^{\sigma} (t) = 0 \, ,
\label{app-ss-meqcombined}
\end{eqnarray}
where $\sigma=0$ corresponds to the scalar and $\sigma = \pm \frac{1}{2}$ to the spinor case. Changing variables
from $t$ to $z= -\e^{2\frac{t}{\tau}}$, and multiplying the left-hand side by a factor $\frac{\tau^2}{4}\frac{1-z}{z}$, we get
\bear
\biggl\lbrack
z(1-z) \frac{d^2}{dz^2} +(1-z)\frac{d}{dz} - \frac{\tau^2\omega_f^2}{4} + \frac{\tau^2\omega_i^2}{4} \frac{1}{z}
+ \Bigl(q^2 E_0^2 \tau^4 + 2i\sigma q E_0\tau^2 \Bigr)\frac{1}{1-z}
\biggr\rbrack
\psi^{\sigma} = 0 \, .
\nonumber\\
\ear
Our aim is to transform this equation into Euler's equation for the hypergeometric function
$\2F1 (\alpha,\beta;\gamma;z)$,
\bear
\biggl\lbrace
z(1-z) \frac{d^2}{dz^2} + \bigl\lbrack \gamma - (\alpha + \beta + 1) z \bigr\rbrack \frac{d}{dz} - \alpha\beta
\biggr\rbrace
\2F1 (\alpha,\beta;\gamma;z) = 0 \, .
\label{eqeuler}
\ear
We can achieve this with the ansatz
\bear
\psi^{\sigma} = z^a (1-z)^b \2F1
\label{ansatz}
\ear
and determining the exponent $a(b)$ by requiring the removal of the poles in $\frac{1}{z}$($\frac{1}{1-z}$). This yields the
two equations
\bear
0 &=& a^2 +  \frac{\tau^2\omega_i^2}{4} \, ,  \label{eqa}\\
0 &=& b(b-1) + q^2 E_0^2 \tau^4 + 2i\sigma q E_0 \tau^2 \, , \label{eqb}
\ear
with solutions
\bear
a_\pm &=& \pm \frac{i}{2}\tau\omega_i \, , \label{sola}\\
b_\pm &=& \frac{1}{2} \mp \sigma \pm i \lambda^{\sigma} \, , \label{solb}
\ear
(see below for the choice of signs)
where we have further introduced
\bear
\lambda^{\sigma} \equiv \sqrt{(qE_0\tau^2)^2-\frac{1}{4} +\sigma^2} \, .
\label{deflambda}
\ear
Assuming that these two conditions are fulfilled, we are down to \eqref{eqeuler} with the identifications
\bear
\gamma &=& 1+ 2a \, , \\
\alpha + \beta &=& 2(a+b) \, , \\
\alpha\beta &=& (a+b)^2 + \frac{\tau^2\omega_f^2}{4} \, .
\ear
Solving this for $\alpha$ and $\beta$ we obtain
\bear
\alpha_\pm &=& a+b \pm \frac{i}{2}\tau \omega_f \, ,\label{alpha}\\
\beta_\pm &=& \alpha_\mp \, .\label{beta}
\ear
It remains to fix the various signs, and to assure the correct normalization. The limit $z\to 0$
has to reproduce the asymptotic initial condition \eqref{condin}. Since
$\2F1(\alpha,\beta;\gamma;0)=1$ we see that we get the correct asymptotic behaviour by
taking the negative sign in the formula for $a$ \eqref{sola}, and that we must also divide by a global factor of
$\sqrt{2\omega_i \e^{\pi\tau\omega_i}}$. To fix the sign in the formulas for $b$ \eqref{solb}, we can use the
fact that in the absence of the external field the solution must remain the initial plane wave \eqref{condin} at all
times. It is easy to see that this is the case only if $b_+$ is chosen for $\sigma=0,\frac{1}{2}$ and $b_-$ for
$\sigma = -\frac{1}{2}$. Finally, the choice of sign in \eqref{alpha}, \eqref{beta} is arbitrary since $\2F1$ is symmetric
in its first two arguments. Thus we choose the upper signs as a convention, and the final result becomes
%
\bear
\psi^0 &=& \frac{1}{\sqrt{2\omega_i \e^{\pi\tau\omega_i}}}
\, z^{a_-}(1-z)^{b_+} \2F1(\alpha_+,\alpha_-,\gamma;z)
\nonumber\\
&=&
\frac{1}{\sqrt{2\omega_i }}
\, \e^{-i\omega_i t}(1+\e^{2\frac{t}{\tau}})^{\half+i\lambda^0}
\nonumber\\
&&\times
\2F1\Bigl(\half+i\lambda^0-\frac{i}{2}\tau(\omega_i-\omega_f),
\half+i\lambda^0-\frac{i}{2}\tau(\omega_i+\omega_f),
1-i\tau\omega_i;-\e^{2\frac{t}{\tau}}\Bigr)
\, ,
\nonumber\\
\psi^{\pm} &=&
\sqrt{\frac{\omega_i \pm k_{\parallel}}{2\omega_i }}
\, \e^{-i\omega_i t}(1+\e^{2\frac{t}{\tau}})^{\frac{1}{2} \mp \half \pm iqE_0\tau^2 }
\nonumber\\
&&\times
\2F1\Bigl(\half\mp \half \pm iqE_0\tau^2
-\frac{i}{2}\tau(\omega_i-\omega_f),
\half\mp \half \pm iqE_0\tau^2
-\frac{i}{2}\tau(\omega_i+\omega_f),
1-i\tau\omega_i;-\e^{2\frac{t}{\tau}}\Bigr) \, .
\nonumber\\
\ear
This is in agreement with \cite{narnik1970,amhuni,Kim:2008yt}.

Thus for the solutions of the corresponding Gelfand-Dikii equations we have
\bear
G^0 &=& \frac{1+\e^{2\frac{t}{\tau}}} {2\omega_i}
\Bigl\vert
\2F1\Bigl(\half+i\lambda^0-\frac{i}{2}\tau \omega_-,
\half+i\lambda^0-\frac{i}{2}\tau\omega_+,
1-i\tau\omega_i;-\e^{2\frac{t}{\tau}}\Bigr)
\Bigr\vert^2
\, ,
\nonumber\\
G^{\pm}  &=& \frac{\omega_i+k_{\parallel}} {2\omega_i}
\Bigl\vert
\2F1\Bigl(\half\mp \half \pm iqE_0\tau^2
-\frac{i}{2}\tau\omega_-,
\half\mp \half \pm iqE_0\tau^2
-\frac{i}{2}\tau\omega_+,
1-i\tau\omega_i;-\e^{2\frac{t}{\tau}}\Bigr)
\Bigr\vert^2
\,,
\nonumber\\
\label{Gsauter}
\ear
denoting now $\omega_\pm \equiv \omega_i \pm \omega_f$.


\subsection{Solution of the scalar mode equation for the solitonic fields}
\label{app-solsol}

The solitonic fields defined by \eqref{Anscal} lead, for $k_{\parallel}= \tilde k_{\parallel}$,
to the scalar mode equation
\begin{equation}\label{sceq}
\ddot{\phi}_{(p)}+\omega_{(p)}^2(t)\phi_{(p)} =0 \, ,
\end{equation}
with frequencies
\bear
\omega_{(p)}^2 (t) = \omega_0^2\Bigl(1 + \frac{p(p+1)}{\cosh^2 (\omega_0 t)}\Bigr) \, .
\ear
The solution parallels the Sauter case.
Changing variables from $t$ to $z = -e^{2\omega_0 t}$,
and multiplying by a factor of $\frac{1}{4\omega_0^2}\frac{1-z}{z}$,
the mode equation becomes
\bear
\biggl\lbrace
z(1-z) \frac{d^2}{dz^2} + \bigl(1-z\bigr) \frac{d}{dz}
+\frac{1-z}{4z} - \frac{p(p+1)}{1-z}
\biggr\rbrace
\phi_{(p)} = 0 \, .
\ear
We apply the substitution 
\bear
\phi_{(p)} = z^a(1-z)^b \2F1(\alpha,\beta;\gamma;z) \, ,
\label{transformtoeuler}
\ear
where removal of the poles fixes
\bear
a_\pm &=& \pm \frac{i}{2} \, , \label{asoliton} \\
b_\pm &=& \half \pm \Bigl(p + \half\Bigr) \, .
\ear
This leads to Euler's equation \eqref{eqeuler} with parameters
\bear
\alpha_\pm &=& a+b \pm a \, , \\
\beta_\pm &=& \alpha_\mp \, , \\
\gamma &=& 1+2a \, .
\ear
The initial condition \eqref{condin} fixes the lower sign for $a_\pm$, and the global normalization.
Requiring that the solution remain the initial plane wave for $p=0$ fixes the upper sign for $b_\pm$.
The final result becomes
\bear
\phi_{(p)} &=& \frac{1}{\sqrt{2\omega_0\e^{\pi}}}\,z^{a_-}(1-z)^{b_+} \2F1 (\alpha_+,\alpha_-,\gamma;z) \nonumber\\
&=&
\frac{1}{\sqrt{2\omega_0}}\,\e^{-i\omega_0t}(1+\e^{2\omega_0 t})^{p+1} \2F1 (p+1-i,p+1,1-i;-\e^{2\omega_0 t})
\, .
\label{solsol}
\ear

\subsection{Solution of the spinor mode equation for the alternative solitonic field}

With the alternative definition \eqref{eA2} of the solitonic fields, the fermionic mode equation \eqref{meqspin} leads to the
Schr\"odinger equation \eqref{sch} with $E=1$ and the potential
\begin{equation}
V(t)=b^2\mathrm{sech}^2 t +b\,\mathrm{sech}\, t \tanh t \, ,
\end{equation}
where we have abbreviated $b \equiv -i \sqrt{p(p+1)}$.
We will solve this equation and also derive the Bogoliubov coefficients, essentially following \cite{khare_jpa1988}.
Note that  \eqref{eA2} also implies the vanishing of
$k_{\parallel}-qA^{\rm new}_{\parallel(p)}(-\infty)$ for $\bk = \tilde {\bf k}$,
which simplifies the asymptotic normalization condition \eqref{condinferm}.

By changing the variable $y=\sinh t$ we cast the Schr\"odinger equation into the form
\begin{equation}\label{eqpsi}
(1+y^2)\psi''(y)+y\psi'(y)+\left[E-\frac{b^2}{1+y^2}-\frac{by}{1+y^2}\right]\psi(y)=0 \, .
\end{equation}
Now let us substitute $\psi(y)=\exp(-b\arctan y) u(y)$ and $(1+iy)/2=z$. Then for $u(z)$ we obtain the hypergeometric equation
\begin{equation}
z(1-z)u''(z)+\left(\frac{1}{2}+ib-z\right)u'(z)-Eu(z)=0 \, .
\end{equation}
Finally the solution of the equation (\ref{eqpsi}) reads (in the following we set $E=1$)
\bear
\psi(y)&=&C_1\,\,_2F_1\Big[-i,i,\half+i b,\half(1+i y)\Big]+C_2\,\,_2F_1 \Big[\half-i-ib,\half+i-ib,\frac{3}{2}-ib,\half(1+iy)\Big] \, .
\label{psinewsol}
\nonumber\\
\ear
Using \cite{abrasteg-book}
\begin{equation}
\lim\limits_{z\to\infty}{}_2F_1\left(a,b,c,z\right)=\frac{\pi}{\sin(b-a)\pi}\left[\frac{(-z)^{-a}}{\Gamma(b)\Gamma(c-a)}-\frac{(-z)^{-b}}{\Gamma(a)\Gamma(c-b)}\right],\quad |\mathrm{Arg}\ z|<\pi
\end{equation}
and $-iy\to \frac{e^{\mp i\pi/2}}{2}e^{\pm t}$ for $t\to\pm\infty$, let us calculate the limits of the solution at $t\to\pm\infty$. They can be represented as
\begin{equation}
\psi(t\to\pm\infty)\to A_{\pm\infty} e^{-it}+B_{\pm\infty} e^{it} \, .
\end{equation}
By assumption,
in the initial state $t\to-\infty$ there is only the positive frequency solution $e^{-it}$, and therefore by \eqref{condinferm}
we must specify $A_{-\infty}=\frac{1}{\sqrt{2}}$ and $B_{-\infty}=0$. This provides
\bear
C_1&=&\frac{\sqrt{\pi}e^{\frac{\pi}{2}(1-b)}}{2^{\half+i}}\,\frac{\Gamma(\half+i+ib)}{\Gamma(\half+i)\Gamma(\half+i b)}\,\Big({\rm sech}\,t+i\tanh t\Big)^{ib}\,\Big(1+\tanh(\pi b)\tanh \pi\Big) \, , \non
C_2&=&-\frac{\sqrt{\pi}\pi\, e^{\frac{\pi}{2}(1+b)}}{2^{1+i-ib}}\,\frac{{\rm sech}\,(b\pi)({\rm sech}\,t)^{ib}\,({\rm sech} \,\pi)(1+i\sinh t)^\half}{\Gamma(\half+i)\Gamma(\half-i+ib)\Gamma(\frac{3}{2}-ib)}\, .\non
\ear
For $A_{\infty}$ and $B_{\infty}$ we get
\begin{equation}
A_{\infty}=\frac{2^{2i}\pi}{\cosh\pi \,\Gamma\left(\frac{1}{2}+ib-i\right)\Gamma\left(\frac{1}{2}-ib-i\right)}\frac{\Gamma(1+i)}{\Gamma(1-i)},\quad B_{\infty}=-\frac{\sinh (\pi b)}{\cosh\pi } \, .
\end{equation}
Thus $A_{\infty}$ and $B_\infty$ are the Bogoliubov coefficients and $|B_\infty|^2$ determines the pair creation probability.
If $b$ is imaginary, as it is in the fermionic case, then
\begin{equation}\label{bcferm}
|A_\infty|^2=\cos^2(\pi|b|)+\sin^2(\pi|b|)\tanh^2\pi ,\quad |B_\infty|^2=\frac{\sin^2(\pi|b|)}{\cosh^2\pi } \, ,
\end{equation}
and $|A_\infty|^2+|B_\infty|^2=1$ corresponds to the relation between Bogoliubov coefficients for fermions.

It is interesting to note that for real $b$ (\ref{eqpsi}) is the usual Schr\"odinger equation for a scalar particle in a Hermitian potential, then
\begin{equation}
|A_\infty^{(s)}|^2=\frac{\cosh (\pi(b-1))\cosh (\pi(b+1))}{\cosh^2\pi },\quad |B_\infty^{(s)}|^2=\frac{\sinh^2(\pi b)}{\cosh^2\pi } \, ,
\end{equation}
and $|A_\infty^{(s)}|^2-|B_\infty^{(s)}|^2=1$ as it should  be in the scalar case.

\end{document}